\documentstyle[12pt]{article}
\input epsf
\def\beq{\begin{equation}}
\def\eeq{\end{equation}}
\def\bea{\begin{eqnarray}}
\def\eea{\end{eqnarray}}

\def\ba{\begin{array}}
\def\ea{\end{array}}
\def\bce{\begin{center}}
\def\ece{\end{center}}

\def\nonu{\nonumber}

\def\Ga{\Gamma}
\def\R{{\bf R}}
\def\Z{{\bf Z}}
\def\C{{\bf C}}
\def\de{\delta}

\def\ze{\zeta}
\def\et{\eta}

\def\La{\Lambda}

\def\Ph{\Phi}

\def\drawbox#1#2{\hrule height#2pt
        \hbox{\vrule width#2pt height#1pt \kern#1pt
              \vrule width#2pt}
              \hrule height#2pt}
\def\Fund#1#2{\vcenter{\vbox{\drawbox{#1}{#2}}}}
\def\fund{\Fund{6.5}{0.4}}
\def\antifund{\bar{\fund}}

\begin{document}
\begin{titlepage}
\rightline{SNUTP-99-011}
\rightline{hep-th/9903181}
\def\today{\ifcase\month\or
January\or February\or March\or April\or May\or June\or
July\or August\or September\or October\or November\or December\fi,
\number\year}
\vskip 1cm
\centerline{\Large \bf Branes at $\C^4/\Ga$ Singularity from Toric Geometry}
\vskip 1cm
\centerline{\sc Changhyun Ahn}
\vskip 0.5cm
\centerline{\it Rm NS1-126,}
\centerline{\it Dept. of Physics,}
\centerline{ \it
Kyungpook National University,} 
\centerline{ \it Taegu702-701, Korea}
\centerline{\tt ahn@kyungpook.ac.kr}
\vskip 0.5cm 
\centerline{and} 
\vskip 1cm
\centerline{ \sc Hoil Kim }
\vskip 0.5cm
\centerline{ \it  Topology and Geometry Research Center,}
\centerline{ \it 
Kyungpook National University,} 
\centerline{\it Taegu 702-701, Korea }         
\centerline{\tt hikim@gauss.kyungpook.ac.kr}
\vskip 1cm
\centerline{\sc Abstract}
\vskip 0.2in
We study toric singularities of the form of $\C^4/\Ga$ for finite abelian 
groups $\Ga \subset SU(4)$. In particular, we consider the simplest case 
$\Ga=\Z_2 \times \Z_2 \times \Z_2$ and find explicitly charge matrices for
partial resolutions of this orbifold by extending the method by Morrison and 
Plesser. We obtain three kinds of algebraic equations, 
$z_1 z_2 z_3 z_4=z_5^2, z_1 z_2 z_3=z_4^2 z_5 $ and $z_1 z_2 z_5 = z_3 z_4$ 
where $z_i$'s  parametrize $\C^5$.  
When we put $N$ D1 branes at this singularity, it is known that
the field theory on the 
worldvolume of $N$ D1 branes is T-dual to $2 \times 2 \times 2 $ brane cub 
model. We analyze geometric interpretation for field theory parameters and
moduli space.    

%\leftline{Feb., 1999}
\end{titlepage}
\newpage

%*****************************************************
%*****************************************************
%*****************************************************
\section{Introduction}
\setcounter{equation}{0}

In \cite{mal}  the large $N$ limit of superconformal field theories (SCFT) 
was described by taking the
supergravity limit on anti-de Sitter (AdS) space.
The scaling dimensions of operators of SCFT were obtained from the 
masses of particles in string/M theory \cite{polyakov}. 
In particular, 
${\cal N}=4$ $SU(N)$ super Yang-Mills theory in 4 dimensions is described by
Type IIB string theory on $AdS_5 \times {\bf S}^5$.  
This AdS/CFT correspondence can be tested by studying
the Kaluza-Klein (KK) states of supergravity theory and 
by comparing them with the chiral primary operators
of the SCFT on the boundary. 
There exist also ${\cal N} =2, 1, 0$ conformal theories in 
4 dimensions which have corresponding supergravity description 
on orbifolds \cite{dm,dgm} of $AdS_5 \times {\bf S}^5$ \cite{kachru}.

On the other hand, T duality transforms orbifold singularity into NS5 branes and
D3 branes into D4 branes. Brane box model \cite{hu} is connected by
D3 branes at $\C^3/\Ga$ singularity, through T duality.
By extending this idea to three kinds of NS5 branes, it was found \cite{gu} that D1 branes,
obtained from D4 branes by T duality, at $\C^4/\Ga$ singularities with $\Ga$ 
an abelian subgroup of $SU(4)$ are related to
brane cub model. They discussed their method by comparing the result of \cite{mohri}
where $(0,2)$ gauge theory on the worldvolume of D1 branes at the singular point of
Calabi-Yau fourfold was studied.

In this paper, we generalize, the work of \cite{mp} where D branes are on the 
non orbifold singularities, to the
case of toric singularities of $\C^4/\Ga$ for finite abelian 
groups $\Ga \subset SU(4)$ and we consider the simplest case 
$\Ga=\Z_2 \times \Z_2 \times \Z_2$. One of the singularities classified in
\cite{mp} is so called conifold singularity which plays an important role in the D brane
worldvolume theory. See the relevant papers \cite{uranga,dm1,gns,lopez,unge}.
In section 2, we find explicitly charge matrices for
partial resolutions of this orbifold and obtain three kinds of algebraic equations.
In section 3, we consider $2 \times 2 \times 2$ brane cub model  
which is dual to the field theory on the 
worldvolume of $N$ D1 branes.
In section 4, we study vacuum moduli space of D1 branes which is a toric variety.
Finally in section 5, we will discuss important open problems 
and comment on the future directions.
%*****************************************************
%*****************************************************
%*****************************************************
\section{ Toric Singularities }
\setcounter{equation}{0}

Before going to our present problems, we review the result of \cite{mp}
in our context and see how it appears.   
If we let $\Ga=\Z_n \times \Z_n $ act on $\C^3$ through the generators
\bea
\mbox{diag}(e^{2\pi i/n}, e^{-2\pi i/n}, 1), \;\;\;\;
\mbox{diag}(e^{2\pi i/n}, 1, e^{-2\pi i/n}),
\eea
then 
the quotient singularity $\C^3/\Ga$ is described by the polygons with vertices
$(0, 0), (n, 0)$ and $(0, n)$.
The vectors consist of all integer vectors
$(k, l)$  with $k \geq 0, l \geq 0 $ and $
k + l  \leq n$.
In the simplest case of $\Z_2 \times \Z_2 $, we  label
the vectors as follows \cite{mp}:
\bea
& & V_0=(0, 0), \;\;\; V_1=(2, 0), \;\;\; V_2=(0, 2) \nonu \\
& & W_0=(1, 1), \;\;\; W_1=(0, 1), \;\;\; W_2=(1, 0). 
\eea
Then the charge matrix in terms of homogeneous coordinates
$X_0, X_1, X_2, Y_0, Y_1, Y_2$ becomes
\bea
\left(
\begin{array}{cccccc}
X_0 & X_1 & X_2 & Y_0 & Y_1 & Y_2  \\
1 & 0 & 0 &  1 & -1 & -1 \\
0 & 1 & 0 & -1 & 1 & -1 \\
0 & 0 & 1 & -1 & -1 & 1  
\end{array}
\right).
\eea
The moment map is given by
\bea
|X_0|^2 + |Y_0|^2 - |Y_1|^2 - |Y_2|^2 = \ze_1,  \nonu \\
|X_1|^2 - |Y_0|^2 + |Y_1|^2 - |Y_2|^2 = \ze_2,  \nonu \\
|X_2|^2 - |Y_0|^2 - |Y_1|^2 + |Y_2|^2 = \ze_3.  
\eea
From now on we use slightly different approach unlike as \cite{mp}
because our presentation  will be more 
clear for the higher dimensional generalization and will lead to same 
result near the region of singularities. The vectors are
\bea
& & w_0=(0, 0), \;\;\; v_1=(2, 0), \;\;\; v_2=(0, 2) \nonu \\
& & u_0=(1, 1), \;\;\; w_1=(1, 0), \;\;\; w_2=(0, 1). 
\eea
From this, the charge matrix is given by and we denote its relation to
those of \cite{mp} as follows:
\bea
\left(
\begin{array}{cccccc}
t_0=Y_0 & x_1=X_1 & x_2=X_2 & y_0=X_0 & y_1=Y_2 & y_2=Y_1  \\
\hline
1 & 0 & 0 &  1 & -1 & -1 \\
0 & 1 & 0 & 1 & -2 & 0 \\
0 & 0 & 1 & 1 & 0 & -2  
\end{array}
\right).
\eea
Then
the moment map is 
\bea
|t_0|^2 + |y_0|^2 - |y_1|^2 - |y_2|^2 = \et_1 = \ze_1,  \nonu \\
|x_1|^2 + |y_0|^2 - 2 |y_1|^2  = \et_2 = \ze_1+\ze_2,  \nonu \\
|x_2|^2 + |y_0|^2 - 2 |y_2|^2 = \et_3 = \ze_1 +\ze_3. 
\eea
Now it is easy to see that all four types classified by the variable 
$\ze_i$ \cite{mp} are exactly
the same as those by the variable $\et_i$ here: $\Z_2 \times \Z_2$ orbifold
corresponds to $\et_1=\et_2=\et_3=0$, suspended pinch point to 
$\et_1=\et_2=0$, conifold to $\et_1=0$ and $\Z_2$ orbifold to $\et_2=0$. 
Furthermore, $\et_2=\et_3=0$ leads to $\Z_2 \times \Z_2$ 
orbifold. Note that alternative description for this space were done in \cite{greene,mr}.

Now we move on one higher dimensional case\footnote{ 
A generalization of the classical MaKay correspondence in complex dimension $d \geq 4$
requires the existence of projective crepant desingularizations. In \cite{dhz}, it was proved
that all $(4; 2)$ ( in their notation ) hyper surface singularities admit crepant projective
desingularizations.
We thank D. Dais for pointing out this.
} and
let $\Ga=\Z_n \times \Z_n \times \Z_n $ act on $\C^4$ through 
the generators
\bea
\mbox{diag}(e^{2\pi i/n}, e^{-2\pi i/n}, 1, 1), \;\;\;
\mbox{diag}(e^{2\pi i/n}, 1, e^{-2\pi i/n}, 1) \;\;\; \mbox{and} \;\;\;
\mbox{diag}(e^{2\pi i/n}, 1, 1, e^{-2\pi i/n}),
\eea
then 
the quotient singularity $\C^4/\Ga$ is described by the polygons with vertices
\bea
(0, 0, 0), \;\;\; (n, 0, 0), \;\;\; (0, n, 0) \;\;\; \mbox{and} 
\;\;\; (0, 0, n).
\eea
A toric Gorenstein canonical singularity of complex dimension 4 is a convex
polygon in $\R^3$ whose vertices have integer coordinates. Let all the vectors
in $\R^3$ have integer coordinates and lie in either 
the interior or the boundary
of the polygons.
The set of vectors consist of all integer vectors $(k, l, m)$ 
with $k \geq 0, l \geq 0, m \geq 0$ and $
k + l + m \leq n$.
In the case of $\Z_2 \times \Z_2 \times \Z_2 $, let us label
10 vectors as follows:
\bea
& & w_0=(0, 0, 0), \;\;\; v_1=(2, 0, 0), \;\;\; v_2=(0, 2, 0), \;\;\;
v_3=(0, 0, 2), \nonu \\
& & w_1=(1, 0, 0), \;\;\; w_2=(0, 1, 0), \;\;\; w_3=(0, 0, 1), \nonu \\
& & u_1=(0, 1, 1), \;\;\; u_2=(1, 0, 1), \;\;\; u_3=(1, 1, 0).
\eea
These vectors are drawed in Fig1(a).
The linear relations between these vectors with integer coefficients
$Q^j$ are
\bea
\sum_{j=1}^3 Q^{j+1} v_j + 
\sum_{j=0}^3 Q^{j+4} w_j +\sum_{j=1}^3 Q^{j+7} u_j =0 
\eea
satisfying the condition
\bea
\sum_{j=1}^{10} Q^j =0.
\label{conditions}
\eea
Let $\vec{Q_1}, \cdots, \vec{Q_6}$ be a basis for the set of all such relations
and use the matrix $(Q_i^j)$ to specify the charges in a representation of
$U(1)^6$ on $\C^{10}$ where $6 = \mbox{the number of vectors}- \mbox{
complex dimension}= 10-4$. Then the singular space is the symplectic reduction
$\C^{10} // U(1)^6$.
Partial and complete resolutions of 
this singularity can be obtained by changing
the center of moment map.

\begin{center}
\leavevmode
\epsfxsize=5in
\epsfbox{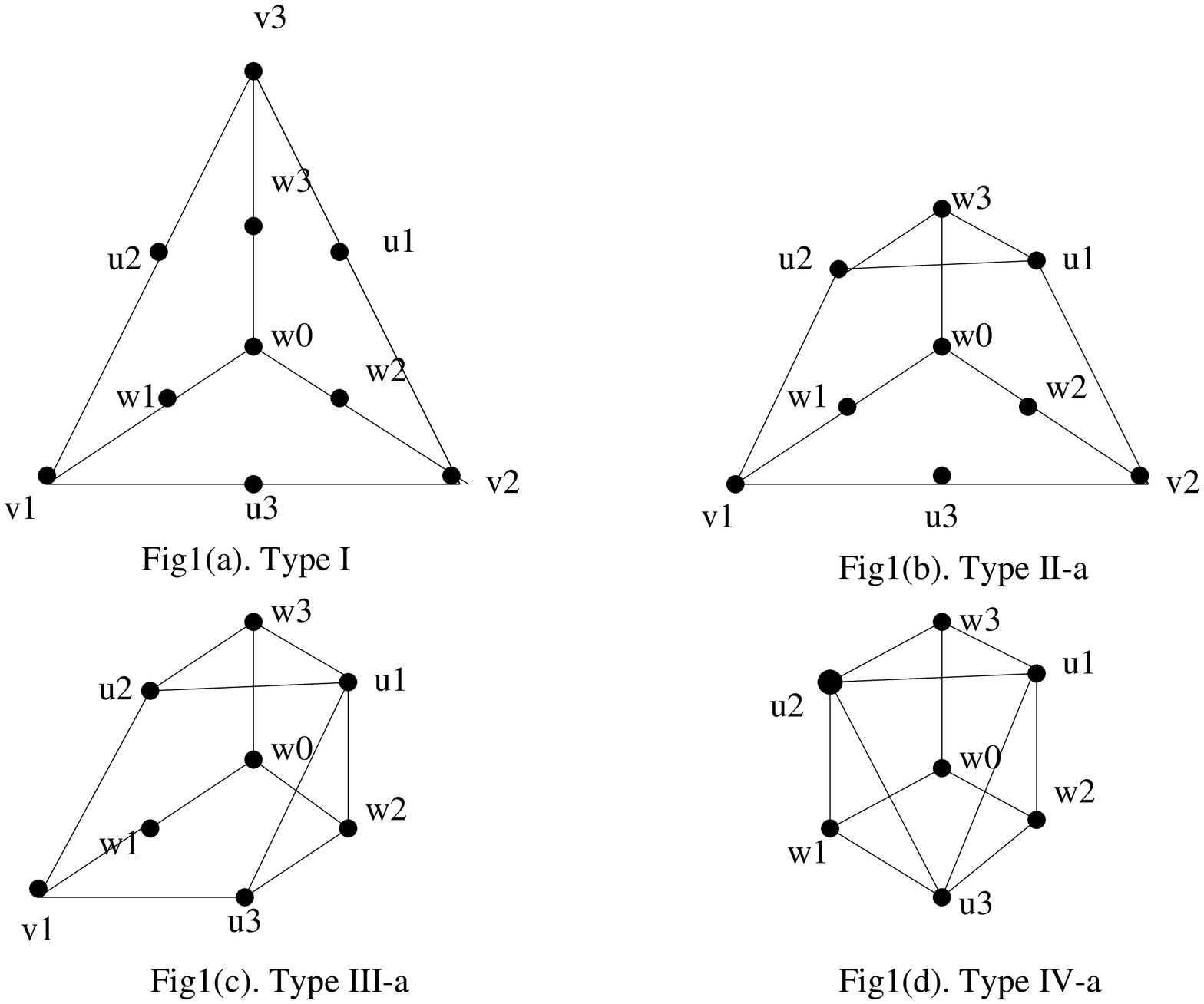}
\end{center}
 
By introducing homogeneous coordinates $t_i, x_i, y_0, y_i( i=1, 2, 3)$
and studying the $U(1)^6$ action, the charge matrix 
is given by
\bea
Q_i^j=
\left(
\begin{array}{cccccccccc}
u_1 & u_2 & u_3 & v_1 & v_2 & v_3 & w_0 & w_1 & w_2 & w_3 \\
\hline
t_1 & t_2 & t_3 & x_1 & x_2 & x_3 & y_0 & y_1 & y_2 & y_3 \\
\hline
1 & 0 & 0 & 0 & 0 & 0 & 1 & 0 & -1 & -1 \\
0 & 1 & 0 & 0 & 0 & 0 & 1 & -1 & 0 & -1 \\
0 & 0 & 1 & 0 & 0 & 0 & 1 & -1 & -1 & 0 \\
0 & 0 & 0 & 1 & 0 & 0 & 1 & -2 & 0 & 0 \\
0 & 0 & 0 & 0 & 1 & 0 & 1 & 0 & -2 & 0 \\
0 & 0 & 0 & 0 & 0 &  1 & 1 & 0 &  0 & -2 
\end{array}
\right).
\label{char}
\eea
Obviously the sum of the elements of each row vanishes according to
(\ref{conditions}). For convenience, we write the 10 vectors in the 1st row and
homogeneous coordinates in the 2nd one.
The moment map is
\bea
|t_1|^2 + |y_0|^2 - |y_2|^2 - |y_3|^2 = \et_1,  \nonu \\
|t_2|^2 + |y_0|^2 - |y_1|^2 - |y_3|^2 = \et_2,  \nonu \\
|t_3|^2 + |y_0|^2 - |y_1|^2 - |y_2|^2 = \et_3,  \nonu \\
|x_1|^2 + |y_0|^2 -2 |y_1|^2  = \et_4,  \nonu \\
|x_2|^2 + |y_0|^2 - 2 |y_2|^2  = \et_5,  \nonu \\
|x_3|^2 + |y_0|^2 - 2 |y_3|^2  = \et_6.
\eea
When $\et_i$'s are generic, the symplectic reduction is smooth.
There are various singularity types for specific values of the $\et_i$'s.
Some of these are illustrated in Fig1(b,c,d).
Each singularity is related to a subgroup $U(1)^k$ of $U(1)^6$ such that
at least $6-k$ of the homogeneous coordinates are uncharged under the
$U(1)^k$. The singular space is described by symplectic reduction of
the space of $\C^{k+4}$ spanned by the remaining $k+4$ homogeneous coordinates
by $U(1)^k$. We list in Table 1. and write the 
condition on $D$-term coefficients and the corresponding charge matrix on the
remaining $k+4$ homogeneous coordinates. By starting from the case $I)$ 
corresponding to zero of all $\et_i$'s and then move on the case in which
five $\et_i$'s are vanishing: There are two cases whether the remaining 
nonzero $\et_j$ are either $\et_1, \et_2, \et_3$ or $\et_4, \et_5, \et_6$. 
Note that
the charge
matrix (\ref{char}) is characterized by two parts: upper  $3 \times 10$ matrix elements
and lower $3 \times 10$ ones. So there are two cases on II): II-a) and II-b). The charge
matrix for II-a) can be read off from I) by removing both 6th row and column.
We can continue all the other cases and end up $VI)$. 
Some of polygons corresponding to
the relevant homogeneous coordinates are given in Fig1.    
As mentioned in the paper of \cite{mp}, in our case also we can consider efficient
description of the singularity possessing fewer fields and smaller group. But we do not
pursue this issue here.

\newpage
\bea
\begin{tabular}[b]{|l|c|}
\hline
I) $\et_i=0( i=1, 2, \cdots, 6 )$ &
$\left(
\begin{array}{cccccccccc}
t_1 & t_2 & t_3 & x_1 & x_2 & x_3 & y_0 & y_1 & y_2 & y_3 \\
1 & 0 & 0 & 0 & 0 & 0 & 1 & 0 & -1 & -1 \\
0 & 1 & 0 & 0 & 0 & 0 & 1 & -1 & 0 & -1 \\
0 & 0 & 1 & 0 & 0 & 0 & 1 & -1 & -1 & 0 \\
0 & 0 & 0 & 1 & 0 & 0 & 1 & -2 & 0 & 0 \\
0 & 0 & 0 & 0 & 1 & 0 & 1 & 0 & -2 & 0 \\
0 & 0 & 0 & 0 & 0 &  1 & 1 & 0 &  0 & -2 
\end{array} \right)$  \\
\hline
II-a)  $\et_i=0( i=1, 2, 3, 4, 5 )$ &
$\left(
\begin{array}{cccccccccc}
t_1 & t_2 & t_3 & x_1 & x_2  & y_0 & y_1 & y_2 & y_3 \\
1 & 0 & 0 & 0 & 0 &  1 & 0 & -1 & -1 \\
0 & 1 & 0 & 0 & 0 &  1 & -1 & 0 & -1 \\
0 & 0 & 1 & 0 & 0 &  1 & -1 & -1 & 0 \\
0 & 0 & 0 & 1 & 0 &  1 & -2 & 0 & 0 \\
0 & 0 & 0 & 0 & 1 &  1 & 0 & -2 & 0 
\end{array} \right)$  \\
\hline
II-b)  $\et_i=0( i=1, 2, 4, 5, 6 )$ &
$\left(
\begin{array}{cccccccccc}
t_1 & t_2 & x_1 & x_2 & x_3  & y_0 & y_1 & y_2 & y_3 \\
1 & 0 & 0 & 0 & 0 &  1 & 0 & -1 & -1 \\
0 & 1 & 0 & 0 & 0 &  1 & -1 & 0 & -1 \\
0 & 0 & 1 & 0 & 0 &  1 & -2 & 0 & 0 \\
0 & 0 & 0 & 1 & 0 &  1 & 0 & -2 & 0 \\
0 & 0 & 0 & 0 & 1 &  1 & 0 & 0 & -2 
\end{array} \right)$  \\
\hline
III-a)  $\et_i=0( i=1, 2, 3, 4 )$ &
$\left(
\begin{array}{cccccccccc}
t_1 & t_2 & t_3 & x_1 & y_0 & y_1 & y_2 & y_3 \\
1 & 0 & 0 & 0 &   1 & 0 & -1 & -1 \\
0 & 1 & 0 & 0 &   1 & -1 & 0 & -1 \\
0 & 0 & 1 & 0 &   1 & -1 & -1 & 0 \\
0 & 0 & 0 & 1 &   1 & -2 & 0 & 0 \\ 
\end{array} \right)$ \\
\hline
\end{tabular}
\eea

Continued

\bea
\begin{tabular}[b]{|l|c|}
\hline
III-b)  $\et_1=\et_4=\et_5=\et_6=0$ &
$\left(
\begin{array}{cccccccccc}
t_1 & x_1 & x_2 & x_3 & y_0 & y_1 & y_2 & y_3 \\
1 & 0 & 0 & 0 &   1 & 0 & -1 & -1 \\
0 & 1 & 0 & 0 &   1 & -2 & 0 & 0 \\
0 & 0 & 1 & 0 &   1 & 0 & -2 & 0 \\
0 & 0 & 0 & 1 &   1 & 0 & 0 & -2 \\ 
\end{array} \right)$ \\
\hline
III-c)  $\et_1=\et_2=\et_4=\et_5=0$ &
$\left(
\begin{array}{cccccccccc}
t_1 & t_2 & x_1 & x_2 & y_0 & y_1 & y_2 & y_3 \\
1 & 0 & 0 & 0 &   1 & 0 & -1 & -1 \\
0 & 1 & 0 & 0 &   1 & -1 & 0 & -1 \\
0 & 0 & 1 & 0 &   1 & -2 & 0 & 0 \\
0 & 0 & 0 & 1 &   1 & 0 & -2 & 0 \\ 
\end{array} \right)$ \\
\hline
IV-a)  $\et_1=\et_2=\et_3=0$ &
$\left(
\begin{array}{cccccccccc}
t_1 & t_2 & t_3 & y_0 & y_1 & y_2 & y_3 \\
1 & 0 & 0 & 1 & 0 & -1 & -1 \\
0 & 1 & 0 & 1 & -1 & 0 & -1 \\
0 & 0 & 1 & 1 & -1 & -1 & 0 \\
\end{array} \right)$ \\
\hline
IV-b)  $\et_1=\et_2=\et_4=0$ &
$\left(
\begin{array}{cccccccccc}
t_1 & t_2 & x_1 & y_0 & y_1 & y_2 & y_3 \\
1 & 0 & 0 & 1 & 0 & -1 & -1 \\
0 & 1 & 0 & 1 & -1 & 0 & -1 \\
0 & 0 & 1 & 1 & -2 & 0 & 0 \\
\end{array} \right)$ \\
\hline
IV-c)  $\et_1=\et_4=\et_5=0$ &
$\left(
\begin{array}{cccccccccc}
t_1 & x_1 & x_2 & y_0 & y_1 & y_2 & y_3 \\
1 & 0 & 0 & 1 & 0 & -1 & -1 \\
0 & 1 & 0 & 1 & -2 & 0 & 0 \\
0 & 0 & 1 & 1 & 0 & -2 & 0 \\
\end{array} \right)$ \\
\hline
\end{tabular}
\eea

Continued

\bea
\begin{tabular}[b]{|l|c|}
\hline
IV-d)  $\et_4=\et_5=\et_6=0$ &
$\left(
\begin{array}{cccccccccc}
x_1 & x_2 & x_3 & y_0 & y_1 & y_2 & y_3 \\
1 & 0 & 0 & 1 & -2 & 0 & 0 \\
0 & 1 & 0 & 1 & 0 & -2 & 0 \\
0 & 0 & 1 & 1 & 0 & 0 & -2 \\
\end{array} \right)$ \\
\hline
V-a)  $\et_1=\et_2=0$ &
$\left(
\begin{array}{cccccccccc}
t_1 & t_2 & y_0 & y_1 & y_2 & y_3 \\
1 & 0 & 1 & 0 & -1 & -1 \\
0 & 1 & 1 & -1 & 0 & -1 \\
\end{array} \right)$ \\
\hline
V-b)  $\et_4=\et_5=0$ &
$\left(
\begin{array}{ccccccccc}
x_1 & x_2 & y_0 & y_1 & y_2  \\
1 & 0 & 1 & -2 & 0  \\
0 & 1 & 1 & 0 & -2  \\
\end{array} \right)$ \\
\hline
V-c)  $\et_1=\et_4=0$ &
$\left(
\begin{array}{cccccccccc}
t_1 & x_1 & y_0 & y_1 & y_2 & y_3 \\
1 & 0 & 1 & 0 & -1 & -1 \\
0 & 1 & 1 & -2 & 0 & 0 \\
\end{array} \right)$ \\
\hline
VI-a)  $\et_1=0$ & 
$\left(
\begin{array}{ccccccccc}
t_1 &  y_0 & y_1 & y_2 & y_3 \\
1 &  1 & 0 & -1 & -1 \\
\end{array} \right) $ \\
\hline
VI-b)  $\et_6=0$ & 
$\left(
\begin{array}{ccc}
x_3 &  y_0 & y_3  \\
1 &  1 & -2    \\
\end{array} \right) $ \\
\hline
\end{tabular}
\eea

Table 1. Charge matrices for partial resolutions of the
$\Z_2 \times \Z_2 \times \Z_2$ orbifold and the conditions on D term 
 coefficients.

For each singularity, we can find $U(1)^k$ invariant monomials
and the equations they satisfy which give the algebraic relation of the 
singularity. They are given in Table 2. 

\newpage
\bea
\begin{tabular}[b]{|l|c|c|}
\hline
I) &
$\begin{array}{ll}
z_1 = y_0^2 y_1 y_2 y_3, &  z_2 = t_2 t_3 x_1^2 y_1  \\ 
z_3 = t_1 t_3 x_2^2 y_2, & z_4 = t_1 t_2 x_3^2 y_3  \\ 
z_5 = t_1 t_2 t_3 x_1 x_2 x_3 y_0 y_1 y_2 y_3 & \empty \\
\end{array}$
& $z_1 z_2 z_3 z_4=z_5^2: (*)$  \\
\hline
II-a)  & 
$\begin{array}{ll}
z_1 = y_0^2 y_1 y_2 y_3, &  z_2 =  t_2 t_3 x_1^2 y_1,  \\ 
z_3 = t_1 t_3 x_2^2 y_2, & z_4 = t_3 x_1 x_2 y_0 y_1 y_2,  \\ 
z_5 = t_1 t_2 y_3 & \empty \\
\end{array}$
& $ z_1 z_2 z_3 = z_4^2 z_5: (**) $  \\
\hline
II-b)  & 
$\begin{array}{ll}
z_1 = y_0^2 y_1 y_2 y_3, &  z_2 =  t_1 t_2 x_3^2 y_3,  \\ 
z_3 = t_1 x_2^2 y_2, & z_4 = t_2 x_1^2 y_1,  \\ 
z_5 = t_1 t_2 x_1 x_2 x_3 y_0 y_1 y_2 y_3 & \empty \\
\end{array}$
& $ z_1 z_2 z_3 z_4 = z_5^2: (*)  $  \\
\hline
III-a)   &
$\begin{array}{ll}
z_1 = y_0^2 y_1 y_2 y_3, &  z_2 = t_2 t_3 x_1^2 y_1,  \\ 
z_3 = t_1 t_2^2 x_1 y_0 y_1 y_3^2, & z_4 = t_3 x_1 y_0 y_1 y_2,  \\ 
z_5 = t_1 t_2 y_3   & \empty \\
\end{array}$
& $z_1 z_2 z_5=z_3 z_4: (***) $  \\ 
\hline
III-b)   &
$\begin{array}{ll}
z_1 = x_1^2 y_1, &  z_2 = y_0^2 y_1 y_2 y_3,  \\ 
z_3 = t_1 x_2^2 y_2, & z_4 = t_1 x_3^2 y_3,  \\ 
z_5 = t_1 x_1 x_2 x_3 y_0 y_1 y_2 y_3   & \empty \\
\end{array}$
& $z_1 z_2 z_3 z_4 =z_5^2: (*) $  \\ 
\hline
III-c)   &
$\begin{array}{ll}
z_1 = y_0^2 y_1 y_2 y_3, &  z_2 = t_2 x_1^2 y_1,    \\ 
z_3 = t_1 x_2^2 y_2, & z_4 = x_1 x_2 y_0 y_1 y_2,  \\ 
z_5 = t_1 t_2 y_3   & \empty \\
\end{array}$
& $z_1 z_2 z_3=z_4^2 z_5: (**) $  \\ 
\hline
IV-a)  & 
$\begin{array}{ll}

z_1 = t_2 y_0 y_1 y_3,  &  z_2 =  t_1 y_0 y_2 y_3,  \\ 
z_3 = y_0^2 y_1 y_2 y_3, & z_4 = t_1 t_2 t_3 y_0 y_1 y_2 y_3, \\
 z_5 = t_3 y_0 y_1 y_2
\end{array}$
& $  z_1 z_2 z_5 =z_3 z_4: (***) $  \\
\hline
IV-b)  & 
$\begin{array}{ll}
z_1 = t_1 y_2, &  z_2 =  y_0^2 y_1 y_2 y_3,  \\ 
z_3 = t_2 x_1^2 y_1, &  z_4 =  x_1 y_0 y_1 y_2,  \\
z_5 = t_1 t_2 y_3 
\end{array}$
& $  z_1 z_2 z_3 =z_4^2 z_5: (**) $  \\
\hline
\end{tabular}
\eea

Continued

\bea
\begin{tabular}[b]{|l|c|c|}
\hline
IV-c)  & 
$\begin{array}{ll}
z_1 = t_1 x_2^2 y_2, &  z_2 =  y_0^2 y_1 y_2 y_3,  \\ 
z_3 = x_1^2 y_1, &  z_4 = x_1 x_2 y_0 y_1 y_2,  \\
z_5 = t_1 y_3
\end{array}$
& $  z_1 z_2 z_3 = z_4^2 z_5: (**) $  \\
\hline
IV-d)  & 
$\begin{array}{ll}
z_1 = x_3^2 y_3, &  z_2 =  y_0^2 y_1 y_2 y_3,  \\ 
z_3 = x_1^2 y_1, & z_4 = x_2^2 y_2,  \\
z_5 = x_1 x_2 x_3 y_0 y_1 y_2 y_3
\end{array}$
& $  z_1 z_2 z_3 z_4 = z_5^2: (*) $  \\
\hline
V-a)   & 
$\begin{array}{ll}
z_1 =  y_0 y_3, &  z_2 = t_1 y_2,  \\ 
z_3 = y_0 y_1 y_2, & z_4 = t_1 t_2 y_3,  \\
z_5 =  t_2 y_1
\end{array}$
& $z_1 z_2 z_5 = z_3 z_4: (***) $  \\
\hline
V-b)   & 
$\begin{array}{ll}
z_1 =  y_0^2 y_1 y_2, &  z_2 = x_1^2 y_1,  \\ 
z_3 =  x_2^2 y_2, & z_4 = x_1 x_2 y_0 y_1 y_2,  \\
\end{array}$
& $z_1 z_2 z_3 = z_4^2 $  \\
\hline
V-c)   & 
$\begin{array}{ll}
z_1 = t_1 y_3, &  z_2 = y_0^2 y_1 y_2 y_3,  \\ 
z_3 =  t_1 y_2, & z_4 = x_1^2 y_1,  \\
z_5 =  t_1 x_1 y_0 y_1 y_2 y_3 
\end{array}$
& $z_1 z_2 z_3 z_4 = z_5^2: (*) $  \\
\hline
VI-a)   & 
$\begin{array}{ll}
z_1 = t_1 y_2 , &  z_2 = t_1 y_3 ,   \\ 
z_3 = y_0 y_2, & z_4 = y_0 y_3 ,  \\ 
\end{array}$
& $z_1 z_4=z_2 z_3 $  \\
\hline
VI-b)   & 
$\begin{array}{ll}
z_1 = x_3^2 y_3, & z_2 = x_3 y_0 y_3,   \\ 
 z_3 = y_0^2 y_3, &   \\ 
\end{array}$
& $ z_1 z_3 = z_2^2 $  \\
\hline
\end{tabular}
\eea

Table 2. Invariant monomials and equations for partial resolutions of the
$\Z_2 \times \Z_2 \times \Z_2$ orbifold. We denote three different kinds of
algebraic equations as $(*), (**)$ and $(***)$ inside the last column.

The presentation of these singularities includes all possible 
toric blown ups by varying the $D$ terms.
The remaining three cases of V-b), VI-a) and VI-b) correspond to
$\Z_2 \times \Z_2$ orbifold, conifold, and $\Z_2$ orbifold, for
the quotient singularity $\C^3/\Ga$  respectively studied in \cite{mp}. 

%*****************************************************
%*****************************************************
%*****************************************************

\section{ Branes At a $\Z_2 \times \Z_2 \times \Z_2$ Singularity }
\setcounter{equation}{0}

Brane configurations \cite{gu} of NS5, NS5', NS5'' and D4 branes 
in Type IIA string theory 
are 
\bea
& & 2 \; \mbox{NS5 branes along} \;\; (x^0, x^1, x^2, x^3, x^4, x^5), 
\nonu \\
& & 2 \; \mbox{NS5' branes along} \;\; (x^0, x^1, x^2, x^3, x^6, x^7), 
\nonu \\
& & 2 \; \mbox{NS5'' branes along} \;\; 
(x^0, x^1, x^4, x^5, x^6, x^7), \nonu \\
& & N \; \mbox{D4 branes along} \;\; (x^0, x^1, x^2, x^4, x^6) 
\eea
where D4 branes are finite in the direction $(x^2, x^4, x^6)$. 
They are bounded in the direction $x^2$
by the NS5'' branes, in the direction $x^4$ by the NS5' branes, and in the
direction $x^6$ by the NS5 branes. The coordinates of 
all branes in $(x^8, x^9)$ should be
equal, 2 NS5 branes should have the same 
position in $x^7$, 2 NS5' branes should
have the same position in $x^5$ and 2 NS5'' branes should have the 
same position
in $x^3$. The low energy effective field theory on the D4 branes is 
$(0,2)$ supersymmetric
 theory in 2 dimension because $(x^0, x^1)$ are the only noncompact directions in
their worldvolume. The existence of each kind of 
NS5 brane breaks one half of the supersymmetries and breaks to 1/8 of
the original supersymmetry. A further half is broken by the presence of
D4 branes and the worldvolume theory therfore is $(0,2)$ supersymmetric
 theory in 2 dimension. The $U(1)_R$ R symmetry of the field theory is 
the rotational symmetry in $(x^8, x^9)$
 directions. Let us perform T duality on the brane cub model along the
 directions $(x^2, x^4, x^6)$.
The T duality along $x^2$ direction transforms the NS5'' branes along $(x^0, x^1,
x^4, x^5, x^6, x^7)$ into 
2 KK'' monopoles.
This gives singularities of type $A_1$. 
The T duality along the $x^4$ direction transforms
NS5' branes along $(x^0, x^1, x^2, x^3, x^6, x^7)$ 
into 2 KK' monopoles which produce type $A_1$ singularity. 
Finally
the T duality along the $x^6$ direction transforms NS5 branes along $(x^0, x^1, x^2, x^3,
x^4, x^5)$into 
2 KK monopoles which
will be $A_1$ singularity. 
Then the final T dual of NS5, NS5' and NS5'' branes is Type IIB string theory
with a complicated geometry in the directions $(x^{2'}, x^3, x^{4'}, x^5, x^{6'}, x^7,
x^8, x^9)$.
Each complex surface of $A_1$ singularity 
is characterized by two 
equations respectively and at the origin all surfaces meet. This can be described by 
a quotient 
singularity of type
$\C^4/\Ga$ with $\Ga=\Z_2 \times \Z_2 \times \Z_2$.
After T duality the original D4 branes become D1 branes located at the singular point.
Since the original D4 branes are bounded by the grid of NS5, NS5' and NS5'' branes,
the T dual D1 branes will be located exactly at the $\C^4/\Ga$ singularity.

%Now we take the $\Z_2\times \Z_2\times 
%\Z_2$ action as follows:
%\bea
%(X, Y, Z, W) \rightarrow (-X, Y, Z, -W), \nonu \\
%(X, Y, Z, W) \rightarrow (X, -Y, Z, -W), \nonu \\
%(X, Y, Z, W) \rightarrow (X, Y, -Z, -W), 
%\eea
The field theory on the worldvolume of $N$ D1 branes on $\C^4$ can be obtained
from ten dimensional ${\cal N}=1 $ $U(N)$ super Yang-Mills by dimensional
reduction. Then the field theory of $D1$ branes on $\C^4/\Ga$ can be obtained by
a projection into $\Ga$ invariant states. The group $\Ga$ acts on the R 
symmetry of the theory and is a subgroup of $SU(4)$. An unbroken $U(1)_R$ R 
symmetry will be present in the quotient theory due to the decomposition
$SU(4) \times U(1)_R \subset SO(8)_R$. The action of $\Ga$ should
be embedded in the Chan-Paton indices of D1 branes. The fields in the resulting
$(0, 2)$ field theory on the D1 branes are those invariant under both 
R symmetry and gauge quantum numbers. 
Following the rule of \cite{dm}, the spectrum can be obtained.
There exist four kinds of complex 
scalar fields $\Ph_{I, I \oplus A_i}, i=1, 2, 3, 4$ according to
the tensor products of the representation ${\bf 4}$ with each irreducible
representation. The fields associated to
the first complex plane ${\Ph}^1_{a, b, c}$ transform in the $(\fund, 
\antifund)$ of $U(N)_{a, b, c} \times U(N)_{a+1, b, c}$. The fields 
${\Ph}^2_{a, b, c}$ transform in the $(\fund, \antifund)$ 
of $U(N)_{a, b, c} \times U(N)_{a, b+1, c}$.
The fields ${\Ph}^3_{a, b, c}$ transform in the $(\fund, \antifund)$ 
of $U(N)_{a, b, c} \times U(N)_{a, b, c+1}$. The last one 
${\Ph}^4_{a, b, c}$ transforms in the $(\fund, \antifund)$ 
of $U(N)_{a, b, c} \times U(N)_{a-1, b-1, c-1}$.  
The chiral multiplets are listed in Table 3 and the corresponding Fermi multiplets
are given in Table 4.

\bea
\begin{tabular}[b]{|c|c|}
\hline
Field & Representations under \nonu \\
& $U(N)_{1,1,1} \times U(N)_{2,1,1} \times U(N)_{1,2,1}
\times U(N)_{2,2,1}$   \nonu \\
 & $ \times U(N)_{1,1,2} \times U(N)_{2,1,2} \times 
U(N)_{1,2,2} \times U(N)_{2,2,2}$ \nonu \\
\hline
$\Ph_{1,1,1}^1; \overline{\Ph_{2,1,1}^1}$ 
& $(\fund, \antifund,1,1,1,1,1,1)$ \nonu \\
$\Ph_{1,2,1}^1;  \overline{\Ph_{2,2,1}^1}$ 
& $(1,1,\fund,\antifund,1,1,1,1)$ \nonu \\
$\Ph_{1,1,2}^1;  \overline{\Ph_{2,1,2}^1}$ 
& $(1,1,1,1,\fund,\antifund,1,1)$ \nonu \\
$\Ph_{1,2,2}^1;  \overline{\Ph_{2,2,2}^1}$ & 
$(1,1,1,1,1,1,\fund, \antifund)$ \nonu \\
\hline
$\Ph_{1,1,1}^2;  \overline{\Ph_{1,2,1}^2}$ & 
$(\fund, 1, \antifund,1,1,1,1,1)$ \nonu \\
$\Ph_{2,1,1}^2;  \overline{\Ph_{2,2,1}^2}$ & 
$(1,\fund,1,\antifund,1,1,1,1)$ \nonu \\
$\Ph_{1,1,2}^2;  \overline{\Ph_{1,2,2}^2}$ & 
$(1,1,1,1,\fund,1,\antifund,1)$ \nonu \\
$\Ph_{2,1,2}^2;  \overline{\Ph_{2,2,2}^2}$ & 
$(1,1,1,1,1,\fund,1, \antifund)$ \nonu \\
\hline  
$\Ph_{1,1,1}^3;  \overline{\Ph_{1,1,2}^3}$ & 
$(\fund,1,1,1, \antifund,1,1,1)$ \nonu \\
$\Ph_{2,1,1}^3;  \overline{\Ph_{2,1,2}^3}$ & 
$(1,\fund,1,1,1,\antifund,1,1)$ \nonu \\
$\Ph_{1,2,1}^3;  \overline{\Ph_{1,2,2}^3} $ &
$(1,1,\fund,1,1,1,\antifund,1)$ \nonu \\
$\Ph_{2,2,1}^3;  \overline{\Ph_{2,2,2}^1} $ &
$(1,1,1,\fund,1,1,1, \antifund)$ \nonu \\
\hline
$\Ph_{1,1,1}^4;  \overline{\Ph_{2,2,2}^4}$ &
$(\fund, 1,1,1,1,1,1,\antifund)$ \nonu \\
$\Ph_{2,1,1}^4;  \overline{\Ph_{1,2,2}^4}$ &
$(1,\fund,1,1,1,1,\antifund,1)$ \nonu \\
$\Ph_{1,2,1}^4;  \overline{\Ph_{2,1,2}^4}$ &
$(1,1,\fund,1,1,\antifund,1,1)$ \nonu \\
$\Ph_{2,2,1}^4;  \overline{\Ph_{1,1,2}^4}$ &
$(1,1,1,\fund, \antifund,1,1,1)$ \\
\hline
\end{tabular}
\eea

Table 3. Chiral multiplets for the theory of D1 branes at singularities. 
Obviously $\Ph_{2,1,1}^1$ is a conjugate representation of $\Ph_{1,1,1}^1$ and
 $\overline{\Ph_{1,1,1}^1}$ is a conjugate representation of 
$\overline{\Ph_{2,1,1}^1}$ and so on.

\bea
\begin{tabular}[b]{|c|c|}
\hline
Field & Representations under \nonu \\
& $U(N)_{1,1,1} \times U(N)_{2,1,1} \times U(N)_{1,2,1}
\times U(N)_{2,2,1}$   \nonu \\
 & $ \times U(N)_{1,1,2} \times U(N)_{2,1,2} \times 
U(N)_{1,2,2} \times U(N)_{2,2,2}$ \nonu \\
\hline
$\La_{1,1,1}^{24}; \overline{\La_{2,1,2}^{24}}$ & 
$(\fund,1,1,1,1,\antifund,1,1)$ \nonu \\
$\La_{2,1,1}^{24};  \overline{\La_{1,1,2}^{24}} 
$ & $(1,\fund,1,1,\antifund,1,1,1)$ \nonu \\
$\La_{1,2,1}^{24};  \overline{\La_{2,2,2}^{24}}$ & 
$(1,1,\fund,1,1,1,1,\antifund)$ \nonu \\
$\La_{2,2,1}^{24};  \overline{\La_{1,2,2}^{24}}$ & 
$(1,1,1,\fund,1,1,\antifund,1)$ \nonu \\
\hline
$\La_{1,1,1}^{14};  \overline{\La_{1,2,2}^{14}}$ & 
$(\fund,1,1,1,1,1, \antifund,1)$ \nonu \\
$\La_{1,2,1}^{14};  \overline{\La_{1,1,2}^{14}}$ & 
$(1,1,\fund,1,\antifund,1,1,1)$ \nonu \\
$\La_{2,1,1}^{14};  \overline{\La_{2,2,2}^{14}}$ & 
$(1,\fund,1,1,1,1,1,\antifund)$ \nonu \\
$\La_{2,2,1}^{14};  \overline{\La_{2,1,2}^{14}}$ & 
$(1,1,1,\fund,1, \antifund,1,1)$ \nonu \\
\hline  
$\La_{1,1,1}^{34};  \overline{\La_{2,2,1}^{34}}$ & 
$(\fund,1,1, \antifund,1,1,1,1)$ \nonu \\
$\La_{2,1,1}^{34};  \overline{\La_{1,2,1}^{34}}$ & 
$(1,\fund,\antifund,1,1,1,1,1)$ \nonu \\
$\La_{1,1,2}^{34};  \overline{\La_{2,2,2}^{34}}$ & 
$(1,1,1,1,\fund,1,1,\antifund)$ \nonu \\
$\La_{2,1,2}^{34};  \overline{\La_{1,2,2}^{34}}$ & 
$(1,1,1,1,1,\fund,\antifund,1)$ \nonu \\
\hline
\end{tabular}
\eea

Table 4. Fermi multiplets for the the theory of D1 branes at singularities.
$\La_{2,1,2}^{24}$ is a conjugate representation of $\La_{1,1,1}^{24}$ and
$\overline{\La_{1,1,1}^{24}}$ is a conjugate representation of $
\overline{\La_{2,1,2}^{
24}}$ and so on. 

The superpotential \cite{mohri} is written, by taking the field $\Ph_{a,b,c}^4$ as special
one, in terms of brane cub model description 
\bea
W & = & \sum_{a,b,c=1}^2 \left[ \La^{24}_{a+1,b,c+1}\left( 
\Ph^1_{a,b,c} \Ph^3_{a+1,b,c}-\Ph^3_{a,b,c} \Ph^1_{a,b,c+1} \right)+
\La^{14}_{a,b+1,c+1} \left( 
\Ph^3_{a,b,c} \Ph^2_{a,b,c+1}-\Ph^2_{a,b,c} \Ph^3_{a,b+1,c} \right) \right.
 \nonu \\
 & & \left. +
\La^{34}_{a+1,b+1,c} \left( 
\Ph^2_{a,b,c}  \Ph^1_{a,b+1,c}- \Ph^1_{a,b,c} \Ph^2_{a+1,b,c} \right) \right].
\label{w}
\eea
Here $\Ph^1_{a,b,c}$ is represented by an arrow which goes from the box $(a, b, c)$ to
the box $(a+1, b, c)$. $\Ph^3_{a+1, b, c}$ is represented by an arrow which goes from
the box $(a+1, b, c)$ to the box $(a+1, b, c+1)$. Furthermore $\La^{24}_{a+1, b, c+1}$ is
represented by an arrow which goes from the box $(a+1, b, c+1)$ to the box $(a+2, b, c+2)$.
So one can see the first term of (\ref{w}) is a gauge invariant object because the final box 
$(a+2, b, c+2)$ is nothing but the initial box $(a, b, c)$ and they form a closed oriented 
triangle. We can check all the other terms in (\ref{w}) form a closed triangles and the 
superpotential $W$ is a gauge
invariant quantity.
The D term equations for this case are
\bea
% |\Phi_{1,1,1}^1|^2 + |\Phi_{1,1,1}^2|^2 + |\Phi_{1,1,1}^3|^2 +
  \sum_{i=1}^4 |\Phi_{1,1,1}^i|^2 - |\Phi_{2,1,1}^1|^2 - |\Phi_{1,2,1}^2|^2- 
  |\Phi_{1,1,2}^3|^2- |\Phi_{2,2,2}^4|^2 & = & r_{1,1,1}, \nonu \\
%  |\Phi_{1,2,1}^1|^2 + |\Phi_{1,2,1}^2|^2 + |\Phi_{1,2,1}^3|^2 +
  \sum_{i=1}^4 |\Phi_{1,2,1}^i|^2 - |\Phi_{2,2,1}^1|^2 - |\Phi_{1,2,1}^2|^2-
  |\Phi_{1,2,2}^3|^2- |\Phi_{2,1,2}^4|^2 & = & r_{1,2,1}, \nonu \\
%  |\Phi_{1,1,2}^1|^2 + |\Phi_{1,1,2}^2|^2 + |\Phi_{1,1,2}^3|^2 +
  \sum_{i=1}^4 |\Phi_{1,1,2}^i|^2 - |\Phi_{2,1,2}^1|^2 - |\Phi_{1,2,2}^2|^2-
  |\Phi_{1,1,1}^3|^2- |\Phi_{2,2,1}^4|^2 & = & r_{1,1,2}, \nonu \\
%  |\Phi_{1,2,2}^1|^2 + |\Phi_{1,2,2}^2|^2 + |\Phi_{1,2,2}^3|^2 +
  \sum_{i=1}^4 |\Phi_{1,2,2}^i|^2 - |\Phi_{2,2,2}^1|^2 - |\Phi_{1,1,2}^2|^2-
  |\Phi_{1,2,1}^3|^2- |\Phi_{2,1,1}^4|^2 & = & r_{1,2,2}, \nonu \\
%  |\Phi_{2,1,1}^1|^2 + |\Phi_{2,1,1}^2|^2 + |\Phi_{2,1,1}^3|^2 +
  \sum_{i=1}^4 |\Phi_{2,1,1}^i|^2 - |\Phi_{1,1,1}^1|^2 - |\Phi_{2,2,1}^2|^2-
  |\Phi_{2,1,2}^3|^2- |\Phi_{1,2,2}^4|^2 & = & r_{2,1,1}, \nonu \\
%  |\Phi_{2,2,1}^1|^2 + |\Phi_{2,2,1}^2|^2 + |\Phi_{2,2,1}^3|^2 +
  \sum_{i=1}^4 |\Phi_{2,2,1}^i|^2 - |\Phi_{1,2,1}^1|^2 - |\Phi_{2,1,1}^2|^2-
  |\Phi_{2,2,2}^3|^2- |\Phi_{2,2,1}^4|^2 & = & r_{2,2,1}, \nonu \\
%  |\Phi_{2,1,2}^1|^2 + |\Phi_{2,1,2}^2|^2 + |\Phi_{2,1,2}^3|^2 +
  \sum_{i=1}^4 |\Phi_{2,1,2}^i|^2 - |\Phi_{1,1,2}^1|^2 - |\Phi_{2,2,2}^2|^2-
  |\Phi_{2,1,1}^3|^2- |\Phi_{2,1,2}^4|^2 & = & r_{2,1,2}, \nonu \\
%  |\Phi_{2,2,2}^1|^2 + |\Phi_{2,2,2}^2|^2 + |\Phi_{2,2,2}^3|^2 +
  \sum_{i=1}^4 |\Phi_{2,2,2}^i|^2 - |\Phi_{1,2,2}^1|^2 - |\Phi_{2,1,2}^2|^2-
  |\Phi_{2,2,1}^3|^2- |\Phi_{1,1,1}^4|^2 & = & r_{2,2,2}
\eea
where
Fayet-Iliopoulos parameter
$r_{a,b,c}=(x^7)_{NS_a} -(x^7)_{NS_{a+1}} +
(x^5)_{NS_b} -(x^5)_{NS_{b+1}}+
(x^3)_{NS_c} -(x^3)_{NS_{c+1}} =
(\de x^7)_a +(\de x^5)_b +(\de x^3)_c$ and satisfies the condition
$\sum_{a,b,c=1}^2 r_{a,b,c} =0$, which is evident from the above
expression, in order to have unbroken supersymmetry.
They also satisfy F term equations given by
\bea
\Ph^1_{a,b,c} \Ph^3_{a+1,b,c} & = & \Ph^3_{a,b,c} \Ph^1_{a,b,c+1}, \nonu \\
\Ph^3_{a,b,c} \Ph^2_{a,b+1,c} & = & \Ph^2_{a,b,c} \Ph^3_{a,b+1,c}, \nonu \\ 
\Ph^2_{a,b,c}  \Ph^1_{a,b+1,c} & = &  \Ph^1_{a,b,c} \Ph^2_{a+1,b,c}, \nonu \\ 
\Ph^4_{a+1,b,c+1} \Ph^2_{a,b-1,c} & = & \Ph^2_{a+1,b,c+1} \Ph^4_{a+1,b+1,
c+1}, \nonu \\
\Ph^4_{a,b+1,c+1} \Ph^1_{a-1,b,c} & = & \Ph^1_{a,b+1,c+1} 
\Ph^4_{a+1,b+1,c+1}, \nonu \\ 
\Ph^4_{a+1,b+1,c}  \Ph^3_{a,b,c-1} & = &  \Ph^3_{a+1,b+1,c} 
\Ph^4_{a+1,b+1,c+1}. 
\label{fterm}
\eea
The first three conditions can be read off from the Fermi superfields in the form of
superpotential in (\ref{w}). The remaining three conditions come from other types of
triangles.

%*****************************************************
%*****************************************************
%*****************************************************
\section{D term Equations and Moduli Space }
\setcounter{equation}{0}

Now we take the $\Z_2\times \Z_2\times 
\Z_2$ action generators $g_1, g_2$ and $g_3$ to act on $\C^4$ as follows:
\bea
g_1: (X, Y, Z, W) \rightarrow (-X, -Y, Z, W), \nonu \\
g_2: (X, Y, Z, W) \rightarrow (-X, Y, -Z, W), \nonu \\
g_3: (X, Y, Z, W) \rightarrow (-X, Y, Z, -W). 
\eea
After diagonalizing, the regular representations of $\Ga$ are given by
\bea
S(g_1) & = & \mbox{diag} ( 1, 1, 1, 1, -1, -1,  -1, -1 ), \nonu \\
S(g_2) & = & \mbox{diag} ( 1, 1,  -1,  -1, 1, 1,  -1, -1 ), \nonu \\
S(g_3) & = & \mbox{diag} ( 1, -1, 1,  -1, 1, -1, 1, -1 ).
\eea
Let $X, Y, Z$ and $W$ be the $8\times8$ matrices from Yang-Mills theory. They
are the lowest components of superfield $\Phi^i_{a,b,c} ( i=1, 2, 3, 4 )$ 
respectively and have the constraints
\bea
& & X=-S(g_1) X {S(g_1)}^{-1}, \;\; X=-S(g_2) X {S(g_2)}^{-1}, 
\;\;  X=-S(g_3) X {S(g_3)}^{-1}, \nonu \\
& & Y=-S(g_1) Y {S(g_1)}^{-1}, \;\; Y=S(g_2) Y {S(g_2)}^{-1}, 
\;\;  Y=S(g_3) Y {S(g_3)}^{-1}, \nonu \\
& & Z=S(g_1) Z {S(g_1)}^{-1}, \;\; Z=-S(g_2) Z {S(g_2)}^{-1}, 
\;\;  Z=S(g_3) Z {S(g_3)}^{-1}, \nonu \\
& & W=S(g_1) W {S(g_1)}^{-1}, \;\; W=S(g_2) W {S(g_2)}^{-1}, 
\;\;  W=-S(g_3) W {S(g_3)}^{-1}. 
\eea
Then the surviving fields after projections are
\bea
 (x_1, x_2, x_3, x_4,x_5,x_6,x_7,x_8) & = &  (X_{18}, X_{27},  X_{36}, X_{45}, X_{54}, X_{63},
X_{72}, X_{81}) \nonu \\
(y_1, y_2, y_3, y_4, y_5, y_6,y_7,y_8) & = &
(Y_{15}, Y_{26}, Y_{37}, Y_{48}, Y_{51}, Y_{62},
Y_{73}, Y_{84}) \nonu \\
(z_1, z_2, z_3, z_4, z_5, z_6, z_7,z_8) & = & (Z_{13},Z_{24},Z_{31},Z_{42},Z_{57},Z_{68},
Z_{75},Z_{86}) \nonu \\
 (w_1, w_2,w_3,w_4,w_5,w_6,w_7,w_8) & = & (W_{12}, W_{21}, W_{34}, W_{43}, W_{56},
W_{65}, W_{78}, W_{87}). 
\label{proj}
\eea
From the F-flatness conditions corresponding to (\ref{fterm})
\bea
& & [X, Y]=0, \;\; [X, Z]=0, \;\; [X, W]=0, \nonu \\
& & [Y, Z]=0, \;\; [Y, W]=0, \;\; [Z, W]=0,  
\eea
the independent variables are as follows
\bea
x_1, x_2, x_3, x_4, x_5, y_1, y_2, y_3, z_1, z_2, w_1.
\eea
That is 11 dimensional affine toric variety. Note that $11-(8-1)=4$.
The cone is generated by the rows of the rectangular $32 \times 11$ matrix.
The remaining variables are expressed in terms of these 11 independent variables.
In usual way, we write them as follows
\bea
\begin{array}{c|ccccccccccc}
& x_1& x_2&x_3 & x_4&x_5 &y_1 &y_2 &y_3 &z_1 &z_2 &w_1  \nonu \\
\hline
\hline
x_1& 1 &0 &0 &0 &0 &0 &0 &0 &0 &0 &0  \\
x_2&  0 & 1&0 &0 &0 &0 &0 &0 &0  &0 &0  \\
x_3&  0 &0 &1 &0 &0 & 0& 0& 0& 0 &0 &0  \\
x_4&  0 &0 &0 &1 &0 &0 & 0& 0& 0 &0 &0  \\
x_5&  0 &0 &0 &0 &1 & 0& 0& 0& 0 &0 & 0 \\
x_6&  0 &0 &-1 &1 &1 &0 &0 &0 & 0 &0 &0  \\
x_7&  0 &-1 & 0& 1 &1 & 0& 0& 0 &0 & 0 & 0 \\
x_8& -1& 0&0 &1 &1 &0 &0 &0 &0  &0 &0  \\
\hline
y_1&  0  &0 &0 &0   & 0  & 1 &0 & 0& 0 & 0&0  \\
y_2&   0 & 0& 0& 0  &0   &0   & 1& 0& 0 &0 & 0 \\
y_3&   0 & 0& 0&0   & 0  & 0  & 0  &1 &0  & 0& 0 \\
y_4& 1&-1 &-1 &1 & 0&-1 &1 &1 & 0 &0 &0  \\
y_5&0 & -1&-1 &1 &1 &-1 &1 & 1& 0 & 0&0  \\
y_6&0 & -1& -1&1 &1 &0 &0 &1 &0  &0 &0  \\
y_7&0 &-1 &-1 &1 &1 &0 & 1&0 & 0 &0 &0  \\
y_8& -1&0 &0 & 0&1 &1 &0 &0 &0  &0 &0 
\end{array}
\eea
Continued

In the above, for example, 6 row implies $x_6 = x_4 x_5 /x_3$ and others hold similarly.

\bea
\begin{array}{c|ccccccccccc}
& x_1& x_2&x_3 & x_4&x_5 &y_1 &y_2 &y_3 &z_1 &z_2 &w_1   \\
\hline
\hline
z_1&0 & 0& 0&0 & 0& 0& 0&0 & 1 &0 &0  \\
z_2&0 & 0&0 &0 &0 &0 &0 &0 & 0 &1 &0  \\
z_3&0 &-1 &0 &1 &0 &-1 & 0&1 & 0 &1 &0  \\
z_4&0 &-1 &0 & 1& 0& -1&0 &1 &1  &0 &0  \\
z_5& 0& 0& 0&0 &0 &-1 &0 &1 &1  &0 &0  \\
z_6& 1&-1 &-1 & 1&0 & -1& 0&1 & 0 &1 & 0 \\
z_7&0 & -1& 0&1 &0 &0 & 0& 0& 0  &1 & 0 \\
z_8& -1&0 &1 &0 &0 &0 &0 &0 & 1 &0 &0  \\
\hline
w_1& 0&0 &0 &0 &0 &0 &0 &0 &0  &0 &1  \\
w_2& 0& 0& -1& 1& 0&-1 &1 &0 &-1  &1 &1  \\
w_3& 0& 0&0 &0 &0 &0 &0 &0 & -1 &1 &1  \\
w_4& 0& 0& -1& 1& 0&-1 &1 &0 & 0 &0 &1  \\
w_5&0 &0 &0 & 0& 0&-1 &1 &0 & 0 &0 &1  \\
w_6&0&0 & -1& 1& 0&0 &0 &0 &-1  &1 & 1 \\
w_7&1 &-1 & -1&1 &0 &-1 &1&0 & -1&1  &1  \\
w_8& -1& 1&0 &0 &0 &0 &0 &0 &0& 0  & 1
\end{array}
\label{data1}
\eea
In order to describe toric variety it is convenient to consider dual cone.
The primitive generators\footnote{The explicit calculation of these is due to
K. Mohri who used  PORTA, c-program of POlyhedron Representation
Transformation Algorithm which can be obtained from the site,
ftp://ftp.zib.de/pub/Packages/mathprog/polyth/index.html. } 
$v_1, \cdots, v_{34} \subset \Z^{11}$ define the linear
map $T: \Z^{34} \rightarrow \bf{N} = \Z^{11}$ which generates the dual cone.
The inner product between 32 vectors in (\ref{data1}) and $v_i(i=1, \cdots, 34)$ are
greater than zero.
\bea
& & v_1=(0, 0, 0, 1, 0, 0, 0, 0, 0, 0, 0), \;\; v_2=(0, 0, 0, 0, 1, 0, 0, 0, 0, 0, 0), \nonu \\
& & v_3=(0, 0, 0, 0, 0, 0, 1, 0, 0, 0, 0), \;\; v_4=(0, 0, 0, 0, 0, 0, 0, 1, 0, 0, 0), \nonu \\
& & v_5=(0, 0, 0, 0, 0, 0, 0, 0, 0, 1, 0), \;\; v_6=(0, 0, 0, 0, 0, 0, 0, 0, 0, 0, 1), \nonu \\
& & v_7=(0, 1, 0, 1, 0, 0, 0, 0, 0, 0, 0), \;\; v_8=(0, 0, 1, 1, 0, 0, 0, 0, 0, 0, 0), \nonu \\
& & v_9=(0, 0, 0, 0, 0, 0, 0, 0, 1, 1, 0), \;\; v_{10}=(0, 0, 0, 0, 0, 0, 0, 0, 1, 0, 1), \nonu \\
& & v_{11}=(0, 0, 0, 0, 0, 1, 1, 1, 0, 0, 0), \;\; v_{12}=(0, 0, 0, 0, 0, 1, 0, 1, 0, 0, 1), \nonu \\
& & v_{13}=(1, 0, 1, 0, 1, 0, 0, 0, 0, 0, 1), \;\; v_{14}=(1, 0, 0, 0, 1, 0, 0, 0, 1, 0, 1), \nonu \\
& & v_{15}=(0, 1, 0, 0, 1, 0, 0, 1, 0, 1, 0), \;\; v_{16}=(0, 0, 1, 0, 1, 0, 1, 0, 0, 1, 0), \nonu \\
& & v_{17}=(0, 0, 1, 0, 1, 0, 0, 1, 0, 0, 1), \;\; v_{18}=(0, 0, 0, 1, 0, 1, 0, 0, 1, 0, 1), \nonu \\
& & v_{19}=(0, 0, 0, 0, 0, 1, 1, 0, 1, 1, 0), \;\; v_{20}=(1, 1, 1, 1, 1, 0, 0, 0, 0, 0, 0), \nonu \\
& & v_{21}=(1, 1, 0, 0, 1, 0, 0, 0, 1, 1, 0), \;\; v_{22}=(1, 0, 0, 1, 0, 1, 0, 0, 1, 0, 1), \nonu \\
& & v_{23}=(0, 1, 1, 1, 0, 0, 1, 1, 0, 0, 0), \;\; v_{24}=(0, 1, 0, 0, 1, 0, 1, 0, 1, 1, 0), \nonu \\
& & v_{25}=(1, 0, 1, 1, 0, 1, 0, 1, 0, 0, 1), \;\; v_{26}=(1, 0, 0, 0, 1, 1, 0, 0, 1, 1, 1), \nonu \\
& & v_{27}=(0, 1, 1, 0, 1, 0, 1, 1, 0, 1, 0), \;\; v_{28}=(1, 1, 1, 1, 0, 1, 1, 1, 0, 0, 0), \nonu \\
& & v_{29}=(1, 1, 1, 0, 1, 0, 1, 1, 0, 1, 0), \;\; v_{30}=(1, 1, 0, 1, 0, 1, 1, 0, 1, 1, 0), \nonu \\
& & v_{31}=(1, 1, 0, 1, 0, 1, 0, 1, 1, 0, 1), \;\; v_{32}=(1, 0, 1, 1, 0, 1, 1, 0, 1, 0, 1), \nonu \\
& & v_{33}=(0, 1, 1, 1, 0, 1, 1, 1, 0, 1, 0), \;\; v_{34}=(1, 1, 1, 0, 1, 1, 1, 1, 1, 1, 1). 
\eea
Also we can get (\ref{data1}) from these primitive generators.
The linear map 
$T$ has the matrix elements $T=(v_1^t, \cdots, v_{34}^t)$ where $t$ is a transpose 
operation.
The kernel of $T$ matrix, which is $23 \times 34$ matrix and denoted by $Q_F$, 
is given by
\bea
Q_F=
\left(
\begin{array}{ccccccccccccc|c}
0&0 &0 &0 &1 &-1 &0 &0 &-1 &1 &0 &0 &0 & \bf{0_{1\times 21}} \\
0&0 &1 &0 &0 &-1 &0 &0 &0 &0 &-1 &1 &0  & \bf{0_{1 \times 21}}  \\
\hline
-1&0 &0 &0 &1 &0 &0 &1 &-1 &0 &0 &0 &-1  &  \\
1&-1 &0 &-1 &-1 &0 &-1 &0 &0 &0 &0 &0 &0 &  \\
1&-1 &-1 &0 &-1 &0 &0 &-1 &0 &0 &0 &0 &0 &  \\
1&-1 &0 &-1 &0 &-1 &0 &-1 &0 &0 &0 &0 &0 &  \\
-1&0 &1 &1 &1 &-1 &0 &0 &-1 &0 &-1 &0 &0 &  \\
0&0 &0 &1 &0 &0 &0 &0 &-1 &0 &-1 &0 &0 &  \\
0&0 &0 &0 &0 &1 &-1 &0 &0 &0 &0 &0 &-1 &  \\
0&0 &0 &0 &0 &1 &-1 &1 &-1 &0 &0 &0 &-1 &  \\
-2&1 &1 &1 &1 &0 &0 &1 &-1 &0 &-1 &0 &-1 & \bf{1_{21\times 21}}  \\
1&0 &-1 &-1 &0 &0 &-1 &-1 &0 &0 &0 &0 &0 &  \\
1&-1 &-1 &0 &0 &0 &-1 &0 &-1 &0 &0 &0 &0 &  \\
-1&1 &1 &0 &0 &0 &0 &0 &0 &0 &-1 &0 &-1 &  \\
-1&0 &1 &1 &0 &0 &0 &1 &-1 &0 &-1 &0 &-1 &  \\
2&-1 &-1 &-1 &-1 &0 &-1 &-1 &0 &0 &0 &0 &0 &  \\
0&1 &0 &0 &0 &1 &-1 &0 &0 &0 &-1 &0 &-1 & \\
1&0 &-1 &-1 &-1 &1 &-1 &0 &0 &0 &0 &0 &-1 &  \\
-1&1 &0 &1 &0 &1 &-1 &1 &-1 &0 &-1 &0 &-1 &  \\
-1&1 &1 &0 &1 &0 &-1 &1 &-1 &0 &-1 &0 &-1 &  \\
-1&1 &0 &1 &1 &0 &0 &0 &-1 &0 &-1 &0 &-1 &  \\
1&0 &0 &0 &-1 &0 &-1 &-1 &0 &0 &-1 &0 &0 & \\
1&0 &0 &0 &0 &0 &-1 &0 &-1 &0 &-1 &0 &-1&
\end{array}
\right).
\eea
Here we denote $1 \times 21$ zero matrix by $\bf{0_{1\times 21}}$ and
$21 \times 21$ identity matrix by $\bf{1_{21 \times 21}}$ for simplicity.
$Q_F$ satisfies $T Q_F^t =0$.
We introduce $11 \times 34$ matrix which satisfies $T U^t =\bf{1_{11 \times 11}}$.
\bea
U=
\left(
\begin{array}{ccccccccccccc|c}
1&0&0&0&0&0&0&0&0&0&0&0&0 &  \\
0&1&0&0&0&0&0&0&0&0&0 & 0 & 0&  \\
0&0&1&0&0&0&0&0&0&0&0 & 0 &0 &  \\
0&0&0&1&0&0&0&0&0&0&0 & 0 & 0&  \\
0&0&0&0&1&0&0&0&0&0&0 & 0 & 0&  \\
0&0&0&0&0&1&0&0&0&0&0 & 0 & 0 & \bf{0_{11\times 21}}  \\
-1&0&0&0&0&0&1&0&0&0&0 &0& 0&   \\
-1&0&0&0&0&0&0&1&0&0&0 &0 & 0&   \\
0&0&0&0&-1&0&0&0&1&0&0 & 0 & 0 &  \\
1&-1&0&0&0&-1&0&-1&0&0& 0& 0& 1 &  \\
0&0&-1&-1&0&0&0&0&0&0& 1&0 & 0&
\end{array}
\right).
\label{U}
\eea
Obviously the choice of $U$ matrix is not unique. For a moment we just put
all other matrix elements from 14 columns to 34 ones.
It is easy to see $U(1)^7$ charge matrix,
\bea
V=
\left(
\begin{array}{c|ccccccccccc}
& x_1& x_2& x_3& x_4& x_5 &y_1 &y_2 &y_3 &z_1 &z_2 &w_1   \\
\hline
q_1& 1&0 &0 &0  &0 &1 &0 & 0& 1& 0&1  \\
q_2& 0& 1&0 &0  &0 &0 &1 &0 &0 &1 &-1  \\
q_3& 0& 0& 1& 0 &0 &0 &0 &1 &-1 &0 & 0 \\
q_4& 0& 0& 0& 1 &-1 &0 &0 &0 &0 &-1 &0  \\
q_5& 0& 0& 0& -1 &1 &-1 &0 &0 &0 &0 &0  \\
q_6& 0& 0& -1& 0 & 0& 0&-1 &0 &0 &0 &0  \\
q_7& 0& -1&0 & 0 &0 &0 &0 &-1 &0 &0 &0 
\end{array}
\right),
\eea
where we can see these charge assignment by remembering (\ref{proj}) .
For example, because of $x_1=X_{18}$, the charges of $x_1$ are $q_1=1$ and
$q_8=-1$ but we took only 7 charges( $q_8=-\sum_{i=1}^7 q_i$ ).
Then
34 homogeneous variables, $p_1, \cdots, p_{34}$ have charges $VU=Q_D$ by 
multiplying $V$ and $U$, given by
\bea
VU=Q_D=
\left(
\begin{array}{ccccccccccccc|c}
1& 0&-1 &-1 &-1 &1 &0 &0 &1 &0 &1 &0&0  &  \\
0& 0&1 &1 &0 &-1 &1 &-1 &0 &0 &-1 & 0&1 &  \\
-1& 0&1 &0 &1 &0 &0 &1 &-1 &0 &0  & 0& 0& \\
-1& 1& 0&1 &-1 &1 &0 & 1& 0& 0& 0 &0 & -1& \bf{0_{7\times 21}} \\
0& 0& 0&-1 & 1& -1& 0&0 &0 &0 &0  &0 &0 &  \\
1& 0& -1& 0& 0& 0&-1 &0 &0 &0 &0  & 0& 0&   \\
1&-1 &0 &0 &0 &0 &0 &-1 &0 &0 &0  &0 & 0& 
\end{array}
\right).
\eea
Combining this with the charge matrix $Q_F$, by including 
Fayet-Illiopoulos parameters in order to describe the moduli space in a simple form, 
 $Q^{\mbox{total}}$
becomes
\bea
Q^{\mbox{total}}=
\left(
\begin{array}{c|c}
 & \xi_1   \\
&  \xi_2  \\
&  \xi_3  \\
Q_D &  \xi_4  \\
&  \xi_5  \\
&  \xi_6  \\
&  \xi_7  \\
\hline
Q_F & \bf{0_{23 \times 1}}
\end{array}
\right)
\eea
where $\xi_i$ correspond to $r_{a,b,c}$ appeared in last section.
In an appropriate region of $\xi$ space, we expect that this will give us exactly same
description of $\Z_2 \times \Z_2 \times \Z_2$ quotient singularity given in (\ref{char}).
The $6\times 10$ matrix elements  can be obtained from $30\times 34$ matrix elements by
doing row operations, the remaining $6 \times 24$ vanish and in the matrix $24 \times
24$ elements 24 homogeneous coordinates can be expressed in terms of 10 independent 
ones.

%*****************************************************
%*****************************************************
%*****************************************************
\section{ Discussion}
\setcounter{equation}{0}

Note that one of the algebraic equations we have found,  $z_1 z_2 z_5 =z_3 z_4$, 
appears 
${\cal N}=1$ $SU(N) \times SU(N) \times SU(N)$ gauge theory on $N$ 
D3 branes at the singular point in 4 dimensions as pointed out by \cite{lopez}. 
The chiral matter fields
transform as
\bea
& & A=(\fund, \antifund, 1), \;\; B=(1, \fund, \antifund), \;\; C=(
\antifund, 1, \fund), \nonu \\
& & \overline{A}=(\antifund, \fund, 1), \;\; \overline{B}=(1, \antifund, \fund), 
\;\; \overline{C}=(
\fund, 1, \antifund). 
\eea 
The minimal set of gauge invariant objects 
\bea
& & z_1 =|A|^2, \;\; z_2=|B|^2, \;\; z_5=|C|^2, \nonu \\
& & z_3 = ABC, \;\; z_4 =\overline{A} \overline{B} \overline{C} 
\eea
satisfy the following constraints
\bea
z_1 z_2 z_5 =z_3 z_4.
\label{constraint}
\eea
This space defines a hypersurface in $\C^5$ and the three lines of singularities meet at the
origin.
In the present our problem, we can think of the following gauge invariant objects
\bea
z_1& = & |\Ph^1_{111}|^2, \nonu \\
z_2 & = & |\Ph^3_{211}|^2, \nonu \\
z_3 & = & \Ph^1_{111} \Ph^3_{211} \La^{24}_{212}, \nonu \\
z_4 & = & \overline{\Ph^1_{111}} \overline{\Ph^3_{211}} \overline{\La^{24}_{212}}, \nonu \\
z_5 & = & |\La^{24}_{212}|^2.
\eea
It is easy to see the gauge invariance by realizing that for the case of the product of two fields
the arrow in brane cub model starts and come back its original position and for the triple
product of fields they form closed oriented triangles like as the one in superpotential.
Of course, they are not independent and are subject to the constraint given by 
(\ref{constraint}).

From 
the toric analysis we have seen previous section, for example,  in codimension 2 in the 
parameter space , we find specific singularity.  
In order to find the worldvolume theory we take $\et_3, \et_4, 
\et_5, \et_6 >> 0$ by keeping
$\et_1=\et_2=0$.
Let us set the vacuum expectation values as follows 
\bea
& & \Ph_{1,1,1}^1   =  \sqrt{\et_3}, \;\;\;\;
 \Ph_{1,2,1}^2   = \sqrt{\et_4} \nonu \\
& &  \Ph_{1,1,2}^3   =  \sqrt{\et_5}, \;\;\;\;
 \Ph_{1,2,2}^4   = \sqrt{\et_6}
\eea
and all others vanish. Then this breaks gauge group down to
$U(N)_{2,1,1} \times U(N)_{2,2,1} \times U(N)_{2,1,2} \times U(N)_{2,2,2}$. 
By inserting these vevs into the superpotential masses for some of other superfields 
appear. They can be integrated out by requiring the equation of motion. We finally get
the light fields and their charges. 

When one is interested in placing M2 branes at the $\C^4/\Ga$
singularity\footnote{ We thank A.M. Uranga for making this paragraph. } in the context of
AdS/CFT correspondence, 
one can choose one compact direction 
in the configuration, shrink it to get
a Type IIA picture and then perform T dualities until the singularities
have been transformed into branes. One possibility is as follows
\bea
& & \mbox{KK monoples }: 
(x^0, x^1, x^2, x^3, x^4, x^5,x^6), \nonu \\ 
& & \mbox{KK' monopoles }: 
(x^0,x^1, x^2, x^3, x^4, x^9, x^ {10}), \nonu \\ 
& & \mbox{KK'' monopoles }: 
(x^0, x^1,x^2, x^5, x^6, x^9, x^{10}), \nonu \\ 
& & \mbox{M2 branes }: ( x^0, x^1, x^2). \nonu
\eea
By
shrinking $x^{10}$ direction, we obtain 
 D6   branes along the directions 
$(x^0, x^1, x^2, x^3, x^4, x^5, x^6)$, 
KK'  monopoles along the directions 
$(x^0, x^1, x^2, x^3, x^4, x^9)$,
KK'' monopoles $(x^0, x^1, x^2, x^5, x^6, x^9)$, and
D2   branes along the directions $(x^0, x^1, x^2)$. 
T duality along $x^4$ and $x^6$ directions lead to
D4   branes along the directions $(x^0, x^1, x^2, x^3, x^5)$,
NS5 branes along the directions $(x^0, x^1, x^2, x^3, x ^4, x^9)$,
NS5' branes along the directions 
$( x^0,  x^1, x^2, x^5, x^6, x^9)$, 
and D4 branes along the directions $(x^0,x^1, x^2, x^4, x^6)$. 
The last three kinds of branes look like a brane box model. 
The first kind of brane (the
D4 which is related to the KK at the beginning) breaks some of the supersymmetry.
%There may be other possibilities to dualize the original singularity
%picture to a brane model. 
Also, the nice symmetry between the different $\Z_k$ factors in the
$\C^4/(\Z_{k} \times \Z_{k'} \times \Z_{k''})$ 
singularity is not clear in these dual pictures
(for instance, above two of the original KK monopoles become NS5 branes,
but the other becomes D4 branes). 
%This could set a limit to the usefulness
%of the picture.
Another compactification \cite{klm} gives 3d ${\cal N}=1$ theory where
all three kinds of NS5 branes have common $x^3$ direction by taking the worldvolume
of third NS5'' branes as $(x^0, x^1, x^2, x^4, x^6, x^8)$ rather than $(x^0, x^1,
x^4, x^5, x^6, x^7)$. This theory is non chiral and needs to be studied further.
Recently \cite{aklm} it was observed that the brane box model should be modified
by adding brane diamond in order to be brane box model as dual of the blowup of the
orbifolded conifold and of the deformed generalized conifold. Although we have not 
thought about this too much we would expect that this phenomena will occur also in the
brane cub model we discussed in section 3.

The toric realization \cite{lv} of ${\bf P}^3$ has a tetrahedron over each 
interior point 
of which there exists 3 torus, which shrinks to a 2 torus at 
four boundary faces, where it shrinks to a circle at each edge of the
tetrahedron, and where it shrinks to a point at each vertex of the 
tetrahedron. Each face of the tetrahedron with a two torus on top 
corresponds to a ${\bf P}^2$ and each edge with a circle on top corresponds 
to a ${\bf P}^1$. However, the ${\bf P}^3$ itself can not be useful, 
it can be used a local
geometry of Calabi-Yau four fold near a singularity. It is known that
$N({\bf P}^3)$ appears part of Calabi-Yau four fold compactification where
the coordinates of which are represented by $(z_1, z_2, z_3, p)$ 
corresponding to ${\bf P}^3$ and the cotangent direction.

When ${\bf P}^3$ is embedded in a Calabi-Yau fourfold, there exists 
a normal direction corresponding to a line bundle on ${\bf P}^3$. The 
property of $c_1=0$ for the fourfold gives that the normal bundle is a 
canonical line bundle. The normal direction to ${\bf P}^3$ can be
identified with the space of $(3,0)$ forms on ${\bf P}^3$. 
Now we have a four dimensional local toric geometry. The extra circle
action plays th role of the rotation on the phase of the normal line
bundle.
The toric realization of $N({\bf P}^3)$ can be viewed as follows:
A copy of ${\bf P}^3$
is placed as the tetrahadron at the bottom. Each semiinfinite face 
emanating from any line on it corresponds to the normal direction of
${\bf P}^3$ in the Calabi-Yau fourfold.  
For M theory on a local singularity of a Calabi-Yau fourfold( 
${\bf P}^3$ shrinking inside a Calabi-Yau fourfold ),
the local model is the canonical bundle over ${\bf P}^3$, $N({\bf P}^3)$.
By modding out $T^3$ action corresponding to three circle action on the 
${\bf P}^3$, there exist 5 dimensional space $N({\bf P}^3)/T^3$ which is
trivial except for the loci where a circle action of $T^3$ has fixed points.
The brane realization of ${\bf P^3}$ was observed in \cite{lv}: 
There exist four faces of the
tetrahedron corresponding to $(p, q, r)$ 4 branes as well as six semiinfinite
faces ending on each six edge of tetrahedron,
corresponding to $(p, q, r)$ external 4 branes. It would be interesting to
study how the toric geometry arise in these brane configuration  in detail.

\vspace{2cm}
\centerline{\bf Acknowledgments} 

We thank A.M. Uranga and K. Mohri for helpful discussions.
This work of CA 
was supported (in part) by the Korea Science and Engineering 
Foundation (KOSEF) through the Center for Theoretical Physics (CTP) at Seoul 
National University. 
 
%****************************************************
%****************************************************
%****************************************************

\end{document}